\documentstyle[preprint,aps,epsf]{revtex}             
\begin{document}
\pagestyle{empty}                                      
\preprint{
\font\fortssbx=cmssbx10 scaled \magstep2
\hbox to \hsize{
\hfill$\raise .5cm\vtop{
\hbox{NHCU-HEP-97-04}\hbox{NCTU-TH-97-03}\hbox{UM-HEP-97-10}}$}
}
\draft
\vfill
\title{QCD Corrections to $b\to s\gamma \gamma $ and Exclusive
$B_s\to \gamma\gamma$
Decay}

\vfill
\author{Chia-Hung V. Chang $^a$, \underline{Guey-Lin Lin}$^b$\thanks{E-mail: glin@beauty.phys.nctu.edu.tw, Tel: 886-3-5731984, Fax: 886-3-5720728.} and York-Peng Yao$^c$}
\address{
$^a$\rm Department of Physics, National Tsing Hua University,
Hsinchu, Taiwan, R.O.C.
}
\address{
$^b$\rm Institute of Physics, National Chiao Tung University,
Hsinchu, Taiwan, R.O.C.
}
\address{$^c$\rm  Randall Laboratory of Physics,  University of
Michigan,
Ann Arbor,
MI 48109, U.S.A.}

%
%
\vfill
\maketitle
\begin{abstract}
The short distance QCD corrections to $b\to s\gamma \gamma$ are calculated
in the leading logarithmic
approximation.  The equivalence of operator basis reduction for S-matrix
elements by using
the equations of motion or by proving a
low energy theorem is discussed. We apply the above results to
the exclusive
$B_s\to \gamma\gamma$  decay. The branching ratio of this decay  is
found to be
$5\times 10^{-7}$ in the Standard Model. We also found that QCD
corrections modify considerably the ratio between CP-even and CP-odd
two-photon amplitudes.
\end{abstract}
%
%
%
%
\pagestyle{plain}


Rare B decays provide useful probes to the flavor structures of the Standard
Model(SM) and its
various extensions. Among them, the radiative decay $B\to X_s
\gamma$\cite{OBSG} has  recently received  the most attention since it
is not only within the reach of current experiments, where CLEO
\cite{CLEO} gives an updated branching ratio
\begin{equation}
Br(B\to X_s \gamma)=(2.32\pm 0.51\pm 0.29\pm 0.32)\times 10^{-4}.
\end{equation}
but also theoretically it has been calculated to the  NLO
accuracy\cite{NBOUND,NANOM}  with the result

\begin{equation}
Br(B\to X_s \gamma)=(3.28\pm 0.33)\times 10^{-4},
\end{equation}
in the SM.
While this decay mode will be continually studied for some time, the
upcoming
B factories at KEK, SLAC and  hadronic B projects at HERA, Tevatron
and LHC offer new opportunities to study many more rare decay
modes.  With these facilities, it will be possible to study decay modes
with branching ratios  as small as 10$^{-8}$.  In this work, we focus
on the exclusive decay
$B_s\to \gamma \gamma$, which, in spite of its small branching ratio, has
a very clear signal where two monochromatic energetic photons are produced
precisely back-to-back in the rest frame of $B_s$.  Within the SM, the electroweak contributions to
this process was calculated without QCD corrections
some time ago\cite{LLY,SW,HK}.  The
branching ratio so obtained was about  $10^{-7}$.
The calculation was later
extended\cite{AT}  to the case of two-Higgs doublet models(2HDM). It
was shown that
$Br(B_s\to \gamma\gamma)$ might be enhanced to $10^{-6}$ in certain
regions of the parameter space. The current
experimental bound on this decay is $Br(B_s\to
\gamma\gamma)<1.48\times 10^{-4}$\cite{L3}.

Before one goes on to study other {\it new physics} which potentially can influence  this
decay, it stands to reason to improve upon previous calculations by incorporating
short distance QCD running from the scale $m_W$ to the scale $m_b$. This will be
achieved by constructing
an effective Hamiltonian for the quark level process $b\to s
\gamma\gamma$ at the scale $\mu=m_b$.  The amplitude for $B_s\to
\gamma\gamma$ can then be determined once a method is devised for
computing the hadronic matrix elements.

At present,
the effective Hamiltonian
$ H_{eff} (b\to s\gamma)$ for $b\to s \gamma$ is already well
understood\cite{REVIEW}.  It is natural to ask what is the
corresponding effective Hamiltonian for
$b\to s\gamma\gamma$. Is it 
identical to  $H_{eff}(b\to s\gamma)$ or
more complicated with extra operators? 
The answer is that, to the order $1/ m_W^2$,
$H_{eff}(b\to s\gamma)$ is sufficient to
generate all contributions to  $b\to s \gamma\gamma$. This can be
understood  by either appplying the
equations of motion \cite{EOM} or  by applying an extension of Low's low
energy theorem
\cite{LLY,LOW}. In the following we shall
illustrate that the above two approaches are in fact equivalent. To
proceed, we write the effective
Hamiltonian for $b\to s \gamma$ as\cite{OBSG}
\begin{equation}
H_{eff}(b\to s \gamma)=-{G_F\over
\sqrt{2}}V^*_{ts}V_{tb}\sum_{i=1}^{8}C_i(\mu)O_i(\mu),
\end{equation}
where
\begin{eqnarray}
O_1 &=& (\bar{s}_ic_j)_{V-A}(\bar{c}_jb_i)_{V-A} \nonumber \\
O_2 &=& (\bar{s}_ic_i)_{V-A}(\bar{c}_jb_j)_{V-A}  \nonumber \\
O_3 &=& (\bar{s}_ib_i)_{V-A}\sum_{q}(\bar{q}_jq_j)_{V-A}  \nonumber
\\
O_4 &=& (\bar{s}_ib_j)_{V-A}\sum_{q}(\bar{q}_jq_i)_{V-A}  \nonumber
\\
O_5 &=& (\bar{s}_ib_i)_{V-A}\sum_{q}(\bar{q}_jq_j)_{V+A}  \nonumber
\\
O_6 &=& (\bar{s}_ib_j)_{V-A}\sum_{q}(\bar{q}_jq_i)_{V+A}  \nonumber
\\
O_7 &=& {e\over
4\pi^2}\bar{s}_i\sigma^{\mu\nu}(m_sP_L+m_bP_R)b_iF_{\mu\nu} \nonumber
\\
O_8 &=& {g\over
4\pi^2}\bar{s}_i\sigma^{\mu\nu}(m_sP_L+m_bP_R)T^a_{ij}
b_jG^a_{\mu\nu} .
\end{eqnarray}
We remark that the decay $b\to s\gamma\gamma$
also receives contributions induced by $\gamma$ and
$Z^0$ penguin diagrams, which are higher-order in electroweak couplings. Effects of such diagrams 
can be summarized into electroweak penguin operators\cite{EWP} which have similar structures as
strong penguins $O_3-O_6$, and an $O(\alpha_{em})$-correction to the Wilson coefficient $C_3$. It is well known that\cite{EWP} $Z^0$ penguin diagram contains the $O(m_t^2)$-contribution which is not much suppressed compared 
to contributions of strong penguins. On the other hand, as we shall see later, the dominant
contributions to $b\to s\gamma\gamma$ actually arise from
$O_1$, $O_2$ and $O_7$. Therefore, in view of simplifying our analysis, we shall 
neglect electroweak penguin contributions.            
     
Concerning the effective Hamiltonian $H_{eff}(b\to s \gamma)$, we note that operator mixings between the set,
$O_1,\cdots, O_6$, and the magnetic penguin $O_7, O_8$
are generally regularization-scheme dependent. However, as pointed
out by Ciuchini {\it et al.} in Ref. [1], the one-loop matrix
elements
of both $b\to s\gamma$ and $b\to s g$ arising from $O_1,\cdots,O_6$
carry compensating
scheme-dependence such that the total physical amplitudes are
independent of regularization
schemes. In view of this, one defines the so called  ``effective
Wilson coefficients"\cite{BMMP},
$C^{eff}_7(\mu)$ and $C^{eff}_8(\mu)$, for operators $O_7$ and
$O_8$, for which
the RG running is scheme independent.
In 't Hooft-Veltman's (HV) scheme for $\gamma_5$, these effective
coefficients  coincide
with original ones. Furthermore, in this scheme, only $O_7$
contributes to the one-loop matrix element of $b\to s \gamma$
\cite{CFMRS}.
Since we shall adopt the HV scheme for subsequent calculations, we  drop
the superscript ``eff" on both $C_7$ and $C_8$ from now on.

To see whether $O_1,\cdots, O_7$ are
sufficient to
generate all contributions to $b\to s \gamma\gamma$, one notes
that operators containing two photon fields are missing from Eq. (4).
Photon fields which constitute operators in this category
can come in  3 ways: from
covariant derivatives $D_{\mu}$,  from field strength tensors $F_{\mu\nu}$
or from some combinations of the two. It has been shown by Grinstein {\it et 
al.}
in Ref. [1] that such operators
can be eliminated or simply reduced to those in Eq. (4) by applying
equations of motion. Therefore,
$H_{eff}(b\to s\gamma)$ is also the effective Hamiltonian for $b\to s
\gamma \gamma$. At this point, we wish to emphasize that  the above
simplifications by using the equations of motion
are applicable only with respect to the $S$ matrix elements \cite{EOM}. The
Green functions depend on the eliminated operators, however.

The above prodecure based on applying
the equations of motion has a close
link with the low energy theorem derived by us some time
ago\cite{LLY}.  Let us illustrate this with the operator
$Q_4\equiv \bar{s}_L{D\!\!\!\!/\ }^3  b_L$\cite{OPER4}.
Upon applying the equation of motion for the b quark, it is easily seen that
$Q_4=im_b^3 \bar{s}_Lb_R$. In this new form, $Q_4$  gives rise to
the flavor changing
self energy. However this contribution will be subtracted out by the
on-shell renormalization
\cite{DE}
\begin{equation}
\bar{s}\Sigma (P^2=m_s^2)=0, \ \ \ \  \Sigma(P^2=m_b^2)b=0.
\end{equation}
Hence, with  equations of motion one concludes that $Q_4$ gives no
contributions to $b\to s \gamma\gamma$. This fact is also realized by
the low energy theorem. There one takes
$Q_4$ as it is initially defined and computes $Q_4$'s contributions to $b-s$ self
energy,  3-point $b\to s\gamma$ and 4-point $b\to s \gamma\gamma$
vertices.  Its total contributions to
$b\to s \gamma\gamma$ are given by summing up the irreducible and
reducible diagrams where the latter come from attaching
$bremsstrahlung$ photons
to the flavor changing self-energy and the three-point  $b\to
s\gamma$ vertex. The renormalization condition,  Eq. (5),
ensures that the end result will be local, and one can further
show that the extraneous contributions cancel out
due to gauge invariance\cite{LLY}. 
Such a
cancellation among reducible and irreducible digrams is what was
referred to as the generalization of  
Low's low energy
theorem\cite{LLY,LOW}.  It is clear that  this theorem is essentially
a verification
of general arguments based upon applying the equations of motion.

With the above issues clarified, we are ready to compute $b\to s
\gamma\gamma$. The amplitude can be separated into irreducible and
reducible parts, as shown in Figs. 1 and 2.
\vskip 1cm
\centerline{Figure 1}
\vskip 1 in
\centerline{Figure 2}
\vskip 1 cm
\noindent We begin with irreducible diagrams which are more
involved. There is only one type of  diagrams which are depicted
generically by Figure 1, where the four-point vertex is a
result of inserting 4-quark operators
$O_1-O_6$. While in the case of $O_1$ and $O_2$, Figure 1 consists of
only one diagram, there are, however,
two different ways of contracting in the case
of $O_3-O_6$, as the light quark
field $q$ appearing in these operators can be external particles as
well. To simplify the algebra,  we  apply Fierz rearrangements to
four-quark operators when necessary. This rearrangement is legitimate
because the $\gamma_5$ we use is simply the product  of
$\gamma^0,\cdots,\gamma^3$.  The
Fierz identities we have used are
\begin{eqnarray}
\left({\bar q}_1 q_2\right)_{V-A} \left({\bar q}_3 q_4 \right)_{V-A}&=&\left({\bar q}_1
q_4\right)_{V-A} \left({\bar q}_3 q_2 \right)_{V-A},
\nonumber \\
\left({\bar q}_1 q_2\right)_{V-A} \left({\bar q}_3 q_4\right)_{V+A}&=&-2\left( {\bar
q}_1(1+\gamma_5) q_4\right) \left({\bar q}_3 (1-\gamma_5)q_2 \right).
\end{eqnarray}
The amplitude for the irreducible diagrams can be written as
\begin{eqnarray}
M_{IR}&=&{16{\sqrt 2}\alpha G_F\over 9\pi}V^*_{ts}V_{tb}\times\nonumber \\
& &\bar{u}(p') \ \left[ \,\sum_q A_qJ(m_q^2)\gamma^{\rho}P_LR_{\mu\nu\rho}
+iB \left(m_sK(m_s^2)P_L+m_bK(m_b^2)P_R\right)T_{\mu\nu} \right. \nonumber \\
& & \left. + C \left(-m_sL(m_s^2)P_L+m_bL(m_b^2)P_R  
\right)\varepsilon_{\mu\nu\alpha\beta}k_1^{\alpha}k_2^{\beta} \, \right] \       
u(p)\epsilon^{\mu}(k_1) \epsilon^{\nu}(k_2),
\end{eqnarray}
where
\begin{eqnarray}
R_{\mu\nu\rho}&=&k_{1,\nu}\varepsilon_{\mu\rho\sigma\lambda}k_1^{\sigma}
k_2^{\lambda}-k_{2,\mu}\varepsilon_{\nu\rho\sigma\lambda}k_1^{\sigma}
k_2^{\lambda}\nonumber \\
&+&k_1\cdot k_2\varepsilon_{\mu\nu\rho\sigma}(k_2-k_1)^{\sigma}
\end{eqnarray}
is a third rank tensor originally constructed by Rosenberg and
Adler\cite{RA}; and
\begin{equation}
T_{\mu\nu}=(k_{2,\mu}k_{1,\nu}-g_{\mu\nu}k_1\cdot k_2).
\end{equation}
Here we adopt the convention: $\varepsilon_{0123}=1$.
The coefficients $A_q$'s, $B$ and $C$ are linear combinations of
Wilson
coefficients and are given by
\begin{eqnarray}
A_u &=&(C_3-C_5)N_c+(C_4-C_6)\nonumber \\
A_d &=&{1\over 4}\left((C_3-C_5)N_c+(C_4-C_6)\right)\nonumber \\
A_c &=&\left((C_1+C_3-C_5)N_c+(C_2+C_4-C_6)\right) \nonumber \\
A_s &=&A_b={1\over 4}\left((C_3+C_4-C_5)N_c+(C_3+C_4-C_6)\right)
\nonumber \\
B &=&C=-{1\over 4}(C_6N_c+C_5).
\end{eqnarray}
Note that the above Wilson coefficients are evaluated at scale
$\mu=m_b$.
Finally the fuctions $J(m^2)$, $K(m^2)$ and $L(m^2)$  are defined by
\begin{eqnarray}
J(m^2)&=&I_{11}(m^2),\nonumber \\
K(m^2)&=&\left(4I_{11}(m^2)-I_{00}(m^2) \right),\nonumber \\
L(m^2)&=&I_{00}(m^2),
\end{eqnarray}
with
\begin{equation}
I_{pq}(m^2)=\int_{0}^{1}{dx}\int_{0}^{1-x}{dy}{x^py^q\over
m^2-2xyk_1\cdot k_2-i\varepsilon}.
\end{equation}

For the reducible diagrams, the one loop contributions with
$O_1$--$O_6$ insertion actually vanish, similar to  what occurs
in $b \to s \gamma $ in HV-scheme. Therefore the amplitude for reducible diagrams is
identical to its lowest
order form\cite{LLY} except for replacing
$C_7(m_W)$ by $C_7(m_b)$.  Hence
\begin{eqnarray}
M_R&=&{{i\sqrt 2}\alpha G_F\over 6\pi}V^*_{ts}V_{tb}C_7(m_b)\times  
\nonumber \\
&&\bar{u}(p') \ \left[  ({1\over p'\cdot k_1}-{1\over p\cdot k_2})
\sigma_{\mu\beta}\sigma_{\nu\alpha}k_1^{\beta}k_2^{\alpha}
+2i ({p'_{\mu}\over p'\cdot k_1}-{p_{\mu}\over p\cdot k_1})
\sigma_{\nu\beta}k_2^{\beta} \right]
\nonumber \\
&&\cdot (m_bP_R+m_sP_L)u(p)\epsilon^{\mu}(k_1)\epsilon^{\nu}(k_2) \   
+ k_1,\mu\longleftrightarrow k_2, \nu.
\end{eqnarray}
From Eqs. (7) and (13), we obtain the total amplitude  for $b\to
s\gamma\gamma$.  

To calculate
$B_s\to \gamma\gamma$, one may follow a perturbative QCD approach
which includes a proof of factorization,
showing that  soft gluon effects can be factorized into $B_s$ meson
wave function;
and a systematic way of resumming large logarithms due to ``hard
gluons" with energies
between $1$ GeV and $m_b$. Since such an approach is technically
rather demanding with uncertain reliability,
we shall adapt
a phenomenological approach where the long distance
effects are  replaced by a few non-perturbative parameters. In
other words we simply evaluate the hadronic matrix element of
$M_{IR}+M_{R}$, relying on a phenomenological model.
Before doing so, it is important to note that $M_R$
is apparently non-local  due
to internal $b$ or $s$ quark propagators.  To handle these non-local
terms, one observes that
the $b$ quark
inside the $B_s$ meson  carries most of the meson energy, and its
four velocity can be treated as equal to that of $B_s$.  Hence one
may write $b$ quark momentum as $p=m_bv$ where
$v$ is the common four velocity of $b$ and $B_s$. With this
parametrization, we have
\begin{eqnarray}
&p\cdot k_1=m_bv\cdot k_1={1\over 2}m_bm_{B_s}=p\cdot k_2,&\nonumber \\
&p'\cdot k_1=(p-k_1-k_2)\cdot k_1=-{1\over2} 
m_{B_s}(m_{B_s}-m_b)=p'\cdot k_2&, 
\end{eqnarray}
where the second equation is based on a constituent picture\cite{LLY} that $b$
and $\bar{s}$ quarks
share the total energy of $B_s$\footnote{Note that the momentum of  $\bar{s}$ quark  is $-p'$ as $p'$ denotes the momentum of $s$ quark in $b\to s \gamma\gamma$.}. Therefore  $-p'$ should be taken as the four
momentum of a constituent $\bar{s}$ quark. With Eq. (14), $M_R$ is
readily made local.
We then compute the amplitude for $B_s\to \gamma\gamma$ using the
following relations
\begin{eqnarray}
\left\langle 0\vert \bar{s}\gamma_{\mu}\gamma_5 b\vert B_s(P)
\right\rangle
&=& -if_{B_s}P_{\mu},\nonumber \\
\left\langle 0\vert \bar{s}\gamma_5 b\vert B_s(P) \right\rangle
&=& if_{B_s}M_B,
\end{eqnarray}
where $f_{B_s}$ is the $B_s$ meson decay constant which is about
$200$ MeV according to
recent Lattice QCD calculations\cite{LATTICE}.

The total amplitude is now separated into a CP-even and a CP-odd part
\begin{equation}
T(B_s\to \gamma\gamma)=M^+F_{\mu\nu}F^{\mu\nu}
+iM^-F_{\mu\nu}\tilde{F}^{\mu\nu}.
\end{equation}
We find that
\begin{equation}
M^+=-{4{\sqrt 2}\alpha G_F\over
9\pi}f_{B_s}m_{B_s}V_{ts}^*V_{tb}\left( Bm_bK(m_b^2)
+{3C_7\over 8\bar{\Lambda} }\right),
\end{equation}
and
\begin{equation}
M^-={4{\sqrt 2}\alpha G_F\over
9\pi}f_{B_s}m_{B_s}V_{ts}^*V_{tb}\left(\sum_q m_{B_s}
A_qJ(m_q^2)+m_bBL(m_b^2)+{3C_7\over 8\bar{\Lambda} }\right),
\end{equation}
where $\bar{\Lambda}=m_{B_s}-m_b$.
The decay width for $B_s\to \gamma\gamma$ is simply
\begin{equation}
\Gamma(B_s\to \gamma\gamma)={m_{B_s}^3\over 16\pi}({\vert M^+\vert
}^2+{\vert M^+\vert }^2).
\end{equation}

To obtain numerical results, we have set light quark masses
to zero and  used \cite{PDG} $m_t=175 \  {\rm GeV}$, $m_b=4.8 \ {\rm
GeV}$
and $m_c=1.5 \ {\rm GeV}$.
Furthermore, we take $m_{B_s}=5.37 \ {\rm GeV}$, $\alpha={1\over
129}$ and  $V_{ts}^*V_{tb}=4\times 10^{-2}$.
The  numerical values for Wilson coefficients $C_1-C_8$ evaluated
at $\mu=m_b$ are listed in Table I.
With the above input parameters, we find $\Gamma( B_s\to
\gamma\gamma)=2.0\times 10^{-10} \ {\rm eV}$ which amounts to a
branching ratio $Br(B_s\to \gamma\gamma)=5.0\times 10^{-7}$, for the given
$\Gamma^{total}_{B_s}=4\times 10^{-4} \  {\rm eV}$. If QCD
corrections are not included, namely
taking $C_2=C_2(m_W)=1$, $C_7=C_7(m_W)$ and setting all the other
Wilson coefficients to zero, one obtains $\Gamma( B_s\to
\gamma\gamma)=1.3\times 10^{-10} \ {\rm eV}$. Therefore
including QCD effects has increased the rate of $B_s\to \gamma\gamma$
by more than $50\%$.

It is interesting  to note that QCD correction modifies contributions
of  irreducible and reducible diagrams
in  opposite ways.  It enhances the contributions of  reducble
diagrams  through the enhancement of
$C_7$, namely $C_7(m_b)/C_7(m_W)\approx 1.5$. Such an enhancement is
already
well known in the decay  $b\to s\gamma$. As a contradistinction, QCD
suppresses the contributions
of irreducible diagrams since its effect essentially replaces
$C_2(m_W)$ with
$N_cC_1(m_b)+C_2(m_b)$\footnote{Since $C_3(m_b),\cdots, C_6(m_b)$ are much smaller
than $C_1(m_b)$
and $C_2(m_b)$, the dominant contribution to $M_{IR}$ is proportional
to $A_c$
which is approximately equal to $N_cC_1(m_b)+C_2(m_b)$.
}. The latter is much smaller than the
former due to
cancellations between $N_cC_1(m_b)$ and  $C_2(m_b)$.
Because QCD gives distinct effects to $M_{IR}$ and $M_R$, the
relative magnitude of
$M^+$ and $M^-$ is also modified accordingly and drastically by QCD.
 From Eqs. (10), (17) , (18) and Table 1,  it is clear that the
magnitudes of $M^+$ and $M^-$ are almost identical, both
dominated by $C_7$
but somewhat corrected by the suppressed
$N_cC_1(m_b)+C_2(m_b)$.
Numerically we have ${\vert M^+\vert}^2/{\vert M^-\vert}^2=0.80$. The
corresponding ratio
without QCD corrections is $0.38$ which is twice smaller.

On the experimental side, as mentioned before,
the future B factories are capable of  probing  B decays with
branching ratio as low as $10^{-8}$.
Therefore one expects the decay $B_s\to \gamma\gamma$ to be seen in
these future facilities.
It is, however, more challenging to separate  partial amplitudes $M^+$
and $M^-$ because it will
require measuring  the spin correlation of  final state photons.

At this point, we wish to state the obvious interdependence between $B_s\to
\gamma\gamma$ and  inclusive
$B\to X_s\gamma$ decays.  It is well known that the later decay
depends on
$C_7(m_W)$ at the tree level, and $C_8(m_W)$ and $C_2(m_W)$ through QCD-induced
operator mixings\cite{OBSG}. CLEO\cite{CLEO} measures
some combinations of these
coefficients, and thus imposes constraints\cite{BMMP,BGGN} on
various extensions of the Standard Model, such as  2HDM
or Minimal Supersymmetric Standard Model(MSSM).
Since $B_s\to \gamma\gamma$ depends on the same set of Wilson
coefficients,
its sensitivity to physics beyond the Standard Model complements
the corresponding sensitivity in $B\to X_s\gamma$\cite{CLY}.

Before closing, we like to comment on the inclusive  $B\to X_s
\gamma\gamma$ decay.
This process has recently been explored without  the inclusion of
QCD corrections\cite{RRS}.
The branching ratio is found to be $10^{-7}$ in the SM.
Experimentally, due to
the appearance of  hadronic final state, it seems to be more difficult
to analyze this process than one can do for $B_s\to \gamma\gamma$. Besides,
photons in this decay may be
soft or collinear which may require extra treatment such as the necessity
of imposing kinematical
cuts. On the theoretical side, there exists a further complication in that photons can be emitted from the spectator quark which forms $B$ with $b$. Such spectator effects, while neglected in Ref. [23], need to be studied carefully in order that additiona

l informations 
on $C_2$ and $C_7$ can be extracted from $B\to X_s \gamma\gamma$.

In conclusion,  we have outlined the equivalence of procedure by using the equations of
motion\cite{EOM} and that by
the low energy theorem\cite{LLY,LOW} to reduce the operator basis, as they are applied  to $b\to
s\gamma\gamma$. We have also
computed the exclusive decay $B_s\to \gamma\gamma$ with the effective
Hamiltonian
$H_{eff}(b\to s\gamma)$. The branching ratio is
found to be $5\times10^{-7}$
which is $50\%$ larger than the result without QCD corrections.
Finally we have argued that one should include spectator effects in the study of inclusive $B\to X_s\gamma\gamma$ decay. 

{\it Note added}. After completing  this work, we became aware of  a
paper by G. Hiller and E. O. Iltan, hep-ph/9704385, where the QCD correction to
$B_s\to \gamma\gamma$  is also discussed. However,  
the crucial contribution from $O_1$ and other contributions from $O_3,\cdots,O_6$ are 
not taken into account in that paper.
\acknowledgments
We thank R. Akhoury for a discussion,
and L. Reina for pointing out a sign error
and a misprint in the earlier version of the manuscript. GLL also likes to thank
particle theory group at U. of
Michigan for their hospitality, where part of this work is done.
The works of  CHC and GLL are supported in part by
National Science Council of R.O.C. under grant numbers  NSC 86-2112-M-007-020
and NSC 86-2112-M-009-012; YPY's work is partially supported
by the U. S. Department of Energy.

\begin{figure}
\caption{The generic irreducible diagram contributing to $b\to
s\gamma\gamma$.
The cross in a circle denotes insertions of operators, $O_1,\cdots,
O_6$. We omit the diagram with two photon lines interchanged.}

\end{figure}

\begin{figure}
\caption{The reducible diagrams contributing to $b\to s\gamma\gamma$
where the interchange
of two photon lines is assumed. The cross in a circle
denotes the insertion of operator $O_7$. }

\end{figure}

\begin{table}
 \caption{Wilson coefficients $C_i(\mu)$ at $\mu=m_b=4.8 \ {\rm GeV}$
in the leading
logarithmic approximations. These values are obtained by taking
$M_W=80.2 \ {\rm GeV}$
and $\alpha_s(M_Z)=0.117$ }
 \begin{tabular}{ccccccccd}
  $C_1(\mu)$     &   $C_2(\mu)$ &  $C_3(\mu)$ & $C_4(\mu)$ &
$C_5(\mu)$ & $C_6(\mu)$ & $C_7(\mu)$ & $C_8(\mu)$    \\ \tableline
  -0.222  &   1.09 & 0.010 & -0.023 &0.007 & -0.028 &   -0.301    &
-0.144     \\
 \end{tabular}
 \label{table1}
 \end{table}


\begin{references}
\bibitem{OBSG}
B. Grinstein, R. Springer and M. B. Wise, Phys. Lett. {\bf B 202} (1988) 138;
Nucl. Phys. {\bf B339} (1990) 269; R. Grigjanis, P. J. O' Donnel, M.
Sutherland and H. Navelet,
Phys. Lett. {\bf B 213} (1988) 355; {\it ibid.} {\bf  B 286} 
(1992) E, 413; M. Misiak, Phys. Lett. {\bf B 269} (1991) 161; K. Adel and Y.-P. Yao, Mod. Phys. Lett. {\bf A 8} (1993) 1679; M.
Ciuchini, E. Franco, G. Martinelli,
L. Reina and L. Silvestrini, Phys. Lett. {\bf B 316 } (1993) 127;
Nucl. Phys, {\bf B 421} (1994) 41;
G. Cella, G. Curci,
G. Ricciardi and A. Vicer\'{e}, Phys. Lett. {\bf B 325} (1994) 227;
Nucl. Phys. {\bf B 431}
(1994) 417; M. Misiak, Nucl. Phys. {\bf B 393} (1993) 23; {\it ibid.} {\bf B 439} (1995) E, 461.
\bibitem{CLEO} {CLEO Collaboration} R. Ammar  {\it et al.},
      Phys. Rev. Lett. {\bf 71} (1993) 674;
      M. S.  Alam  {\it et al.},
       {\it ibid.} {\bf 74} (1995) 2885.
\bibitem{NBOUND}
K. Adel and Y.-P. Yao, Phys. Rev. {\bf D 49} (1994) 4945;
C. Greub and T. Hurth, hep-ph/9703349.
\bibitem{NANOM}
K. G. Chetyrkin, M. Misiak and M. M\"{u}nz, hep-ph/9612313.
\bibitem{LLY}
G.-L. Lin, J. Liu and Y.-P. Yao, Phys. Rev. Lett. {\bf 64}
(1990) 1498; Phys. Rev. {\bf D 42} (1990) 2314.
\bibitem{SW}
H. Simma and D. Wyler, Nucl. Phys. {\bf B 344} (1990) 283.
\bibitem{HK}
S. Herrlich and J. Kalinowski, Nucl. Phys. {\bf B 381} (1992) 501.
\bibitem{AT}
T. M. Aliev and G. Turan, Phys. Rev. {\bf D 48} (1993) 1176.
\bibitem{L3}
M. Acciarri {\it et. al.} (L3 Collaboration), Phys. Lett. {\bf B
363} (1995) 127.
\bibitem{REVIEW}
For a review on earlier literatures, see G. Buchalla, A. J. Buras and
M. E. Lautenbacher, Review of Modern Physics,
{\bf 68} (1996) 1125.
\bibitem{EWP}
J. M. Flynn and L. Randall, Phys. Lett. {\bf B 224} (1989) 221;
A. J. Buras, M. Jamin, M. E. Lautenbacher and P. Weisz, Nucl. Phys. {\bf B 400} (1993) 37;
A. J. Buras, M. Jamin and M. E. Lautenbacher,
{\it ibid.} {\bf B 400} (1993) 75; M.
Ciuchini, E. Franco, G. Martinelli and
L. Reina, Nucl. Phys. {\bf B 415} (1994) 403.  
\bibitem{EOM}
General discussions on the uses of equations of motion in field
theory can be found in
H. D. Politzer, Nucl. Phys. {\bf B 172} (1980) 349; C. Arzt, Phys.
Lett. {\bf B 342} (1995) 189.
For specific discussions with respect to rare B decays, see
H. Simma, Z. Phys. {\bf C 61} (1994) 67.
\bibitem{LOW}
F. E. Low, Phys. Rev. {\bf 110} (1958) 974.
\bibitem{BMMP}
A. J. Buras, M. Misiak, M. M\"{u}nz and S. Porkorski, Nucl. Phys.
{\bf B 424} (1994) 374.
\bibitem{CFMRS} See discussions by Ciuchini {\it et al.} in Ref. [1].
\bibitem{OPER4} This labeling is according to the convention of
Grinstein {\it et al.}
in Ref. [1].
\bibitem{DE}
N. G. Deshpande and G. Eilam, Phys. Rev. {\bf D 26} (1982) 2463.
\bibitem{RA}
L. Rosenberg, Phys. Rev. 129 (1963) 2786;
S. L. Adler, {\it ibid.} {\bf 177} (1969) 2426.
\bibitem{LATTICE}
For a review of recent progress, see J. Flynn, in International
Conference on High
Energy Physics, Warsaw, Poland, 1996.
\bibitem{PDG}
Particle Data Group, R. M. Barnett {\it et. al.} Phys. Rev. {\bf D
54} (1996) 1.
\bibitem{BGGN}
R. Barbieri and G. F. Giudice, Phys. Lett. {\bf B 309} (1993) 86;
R. Garisto and J. N. Ng, {\it ibid.} {\bf B 315} (1993) 372.
\bibitem{CLY}
C.-H. V. Chang, G.-L. Lin and Y.-P. Yao, work in progress.
\bibitem{RRS}
L. Reina, G. Ricciardi and A. Soni, Phys. Lett. {\bf B 396}
(1997) 231.
\end{references}
\end{document}